\documentclass[%
reprint,
superscriptaddress,
amsmath,amssymb,
aps,
]{revtex4-1}

\usepackage{color}
\usepackage{graphicx}
\usepackage{dcolumn}
\usepackage{bm}

\begin {document}

\title{Trace of anomalous diffusion in a biased  quenched trap model
}

\author{Takuma Akimoto}
\email{takuma@rs.tus.ac.jp}
\affiliation{%
  Department of Physics, Tokyo University of Science, Noda, Chiba 278-8510, Japan
}%


\author{Keiji Saito}
\affiliation{%
  Department of Physics, Keio University, Yokohama, 223-8522, Japan
}%


\date{\today}

\begin{abstract}
Diffusion on a quenched heterogeneous environment in the presence of bias is considered analytically. The first-passage-time statistics 
can be applied to obtain the drift and the diffusion coefficient in periodic quenched environments.  We show several transition points at which  sample-to-sample fluctuations of the drift or the diffusion coefficient remain large even when the system size becomes large, i.e., non-self-averaging. Moreover, we find that 
the disorder average of the diffusion coefficient diverges or becomes zero when the corresponding annealed model generates superdiffusion or subdiffusion, respectively. This result implies that  anomalous diffusion in an annealed model is traced by anomaly of the diffusion coefficients in the corresponding quenched model. 
\end{abstract}

\maketitle



\section{Introduction}
Anomalous transport characterized by a nonlinear growth of the mean squared displacement (MSD) or 
the mean displacement (MD) in the presence of bias is a ubiquitous phenomenon in nature. 
Since a discovery of anomalous transport in amorphous materials \cite{Scher1975},  
considerable efforts have been made to unveil anomalous physical features in experiments such as 
biological systems \cite{Caspi2000, Wong2004, Golding2006, Szymanski2009, Bronstein2009, Gal2010, Weigel2011, Jeon2011, Tabei2013, Hofling2013, Manzo2015}. One of physical mechanisms that produce anomalous transport is attributed to  a heterogeneous environment, where 
local diffusivity is spatially heterogeneous or changes with time. 
 
A typical model of anomalous diffusion on heterogeneous environments is a quenched trap model (QTM) \cite{bouchaud90}. 
In this model, the spatial heterogeneity is represented by a quenched random energy landscape. While this model is simple and 
has been investigated for decades, there are few exact results on the QTM \cite{derrida1983velocity}. This is because one has to take into account 
how a random walker visits a site. 
In particular,  one needs to calculate the number of visits to a site by a random walker to obtain an exact result. The MSD of the QTM 
shows anomalous diffusion when the temperature is below the glass temperature, where the mean waiting time diverges \cite{bouchaud90}. 
By a scaling argument, it is known that the power-law exponent of the MSD depends on the spatial dimension \cite{bouchaud90}. 
Furthermore, fluctuations of the diffusion coefficients obtained by single trajectories depend intrinsically on the dimension \cite{Miyaguchi2011, Miyaguchi2015}.

Continuous-time random walk (CTRW) is an annealed model of the QTM, which is widely used to investigate anomalous diffusion because 
its analytical treatment is possible due to the spatial homogeneity \cite{metzler00}. In the CTRW, the waiting-time distribution does not 
depend on the site but is identical for all the sites. This is a significant difference between the QTM and the CTRW, which gives a rich physical 
feature such as sample-to-sample fluctuations \cite{Miyaguchi2011, Miyaguchi2015, Luo2015, Akimoto2016, *Akimoto2018}. However, 
the CTRW becomes a good approximation of the QTM when the spatial dimension is greater than two or in the presence of bias \cite{Machta1985}. 
In these situations, a random walker can visit a new site at constant non-zero probability, which reduces a risk of returning to the sites that 
a random walker visited before. Remarkable features of the CTRW are observed in the time-averaged-base MSD (TAMSD). 
When the mean waiting time diverges, the TAMSD is not coincided with  the ensemble-averaged-base MSD, i.e., ergodicity breaking 
\cite{He2008, Neusius2009, Miyaguchi2013, Metzler2014}. In our previous studies \cite{Akimoto2016, *Akimoto2018}, we showed that 
the QTM with finite system size is ergodic; i.e., the TAMSD is equivalent to the corresponding MSD, and that the TAMSD becomes non-self-averaging 
when the temperature is below the glass temperature. 

Effects of bias in the CTRW have been investigated in the regime where the mean waiting time is finite but the second moment diverges. 
In this regime, the drift is normal but the variance of the displacement (VD) shows superdiffusion, whereas the MSD in the absence of bias 
is normal \cite{Burioni2013, *Burioni2014, Akimoto2018ergodicity, *hou2018biased}. This phenomenon  is called field-induced superdiffusion. 
Field-induced superdiffusion is ubiquitous phenomenon observed in crowded systems such as supercooled liquids \cite{Schroer2013, Benichou2013, benichou2013b, Gradenigo2016, Leitmann2017}. However, field-induced superdiffusion in heterogeneous quenched 
environments has not been studied so far. 

In this paper, we discuss how a bias affects transport properties in the QTM with a finite system size, where we assume a periodic boundary condition. 
In a periodic system, the first-passage-time (FPT) statistics plays an important role in obtaining the transport properties such as drift and diffusivity 
\cite{Reimann2001, *Reimann2002}. Applying the FPT statistics in a biased QTM \cite{Akimoto2019}, we show non-self-averaging properties 
of the MD and the VD. Comparing with the CTRW results, we provide 
an interesting connection between the annealed and quenched models.

\section{model}
Here, we consider a random walk on a one-dimensional random energy landscape, i.e., QTM \cite{bouchaud90}, 
where the energy landscape is quenched, and  assume that the landscape is arranged periodically. 
The probabilities of stepping to the right and the left site are denoted by $p$ and $q=1-p$, respectively. We assume that the lattice constant is unity and that 
 the tops of the potentials are flat; i.e., the tops are the same hight for all sites. In other words, probabilities $p$ and $q$ do not depend on the site. 
In particular, we consider the case of $p\ne \frac{1}{2}$, i.e., a biased QTM, and $p>1/2$ for simplicity. 

Quenched disorder implies that when realizing the random energy landscape it does not change with time.   
The number of  lattice sites with different energies is $L$; i.e., the energy landscape of the system is periodically arranged with period $L$.
At each lattice point, the depth $E>0$ of an energy trap is randomly assigned and quenched. The depths are 
independent identically distributed random (IID) variables with an exponential distribution,
$\rho(E) = T^{-1}_g \exp(-E/T_g)$, where $T_g$ is called a glass temperature. 
A particle can escape from a trap and jump to one of the nearest neighbors. 
Escape times from a trap are IID random variables with an exponential distribution  
and follows the Arrhenius law; i.e., the mean escape time of the $k$th site is given by $ \tau_{k}  = \tau_c \exp(E_{k}/T)$, 
where $E_{k}$ is the depth of the energy at site $k$, 
$T$  the temperature, and $\tau_c$ a typical time. 
The probability that escape time $\tau$ is smaller than $x$ is given by 
$\Pr (\tau \leq x) \cong \Pr (E \leq T \ln (x/\tau_c))$.
Because the probability density function (PDF) of random variable $E$ follows $\rho(E)$, 
the PDF of $\tau$ for infinite systems, denoted by $\psi_\alpha(\tau)$, becomes 
\begin{equation}
\int_\tau^\infty d\tau' \psi_\alpha (\tau') \cong \left(\frac{\tau}{\tau_c} \right)^{-\alpha}\quad (\tau \geq \tau_c)
\label{power-law-pdf}
\end{equation}
with $\alpha \equiv T/T_g$ \cite{Bardou2002, Akimoto2018}. 
 
 The master equation of the biased QTM can be represented by the quenched disorder realization.
Let $P_{k}(t)$ be the probability of finding a particle at site ${k}$ at time $t$. 
The master equation for the $i$th disorder realization, where the mean escape time at site $k$ is denoted by $\tau_{k}^{(i)}$,  
is given by 
\begin{equation}
\frac{dP_{k}(t)}{dt} = p\frac{P_{k-1}(t)}{\tau_{k-1}^{(i)}}  + q\frac{P_{k+1}(t)}{\tau_{k+1}^{(i)}}  - \frac{P_{k}(t)}{\tau_{k}^{(i)}}.
\end{equation}
We consider the periodic boundary condition, i.e.,  $P_0(t)=P_L(t)$, $P_{L+1}(t)=P_1(t)$,  $\tau_0^{(i)}=\tau_L^{(i)}$, and $\tau_{L+1}^{(i)}=\tau_1^{(i)}$. 
It follows  that the steady state can be obtained as
\begin{equation}
P_{k}^{\rm st} = \frac{\tau_{k}^{(i)}}{L \mu_i},
\label{steady_state}
\end{equation}
where $P_{k}^{\rm st}$ is the probability of finding a particle at site $k$ in the steady state and $\mu_i$ is the sample mean 
for the $i$th disorder realization: 
\begin{equation}
\mu_i = \frac{1}{L} \sum_{k=0}^{L-1}\tau_{k}^{(i)} .
\end{equation}
The steady state is exactly the same as the equilibrium state under no bias \cite{Akimoto2016, Akimoto2018}. 
This is because the bias we consider here does not change the shape of the random energy landscape.

\section{First-passage-time statistics and diffusive properties}

In our previous study \cite{Akimoto2019}, we derive the first-passage-time (FPT) statistics for the biased QTM. Here, we review the FPT statistics in the biased 
QTM and show that they play an important role in obtaining the diffusive properties such as drift and diffusivity in the system. 
The FPT in the biased QTM is defined as  the time when a particle starting from the origin reaches site $L$, i.e., the right boundary. 
We note that the target is located at site $L$ only (site $-L$ is not a target) whereas we consider a periodic landscape. 
In the large-$L$ limit,  the mean FPT (MFPT) and the variance of the FPT (VFPT) for  the $i$th disorder realization  are represented by 
\begin{align}
 \langle T \rangle                 & \sim  \frac{ L \mu_i}{p-q} ,  \label{mfpt_qtm} \\ 
 \langle \delta T^2 \rangle   & \sim  \frac{L \{ \sigma_i^2(p-q) + \mu_i^2\}}{(p-q)^3} , 
 \label{var_fpt_qtm}
 \end{align}
where $\delta T \equiv T - \langle T \rangle $ and $\sigma_i^2$ is the sample variance, i.e.,
\begin{equation}
\sigma_i^2 = \frac{1}{L} \sum_{n=0}^{L-1} (\tau_n^{(i)})^2 - \mu_i^2 \, .
\end{equation}
These formulae are exact for any quenched disorder realizations in the large-$L$ limit and  depend crucially on the disorder realization. 
Therefore, sample-to-sample fluctuations for the MFPT and the VFPT  become significant when $\alpha$ is smaller than two   \cite{Akimoto2019}. 

Here, we connect the FPT statistics with the MD and the VD. 
Let $n_t$ be the number of events that a particle crosses the right boundary, i.e., stepping from  $L-1$ to  $L$ site, which is equivalent to make 
a counterclockwise revolution on a ring. 
In the large-$L$ limit, the probability that a particle starting from the origin crosses the left boundary, i.e., stepping from  $-L+1$ to  $-L$ site, 
becomes zero. Therefore, a particle always resides in a interval $[ (n_t -1)L, n_t L)$ in the large-$L$ limit. It follows that 
 the displacement, $\delta x_t \equiv x(t) - x(0)$, can be represented by
\begin{equation}
\delta x_t = L n_t + C_L,
\label{xt_nt}
\end{equation}
where $x(t)$ is a position of a particle at time $t$ and $C_L$ is a residual term that is considered to be a 
random variable whose support is $-L<C_L < L$, and thus $\langle |C_L| \rangle =O(L)$. 
Because time when a particle crosses the right boundary is an IID random variable,  
the process of $n_t$ is described by a renewal process \cite{Cox}.
By the renewal theory \cite{Cox}, the mean of $n_t$ is given by
\begin{equation}
\langle n_t \rangle \sim \frac{t}{\mu}\quad (t\to\infty),
\end{equation} 
where $\mu$ is the mean inter-event time, which is equivalent to MFPT $\langle T \rangle$. 
Therefore, the MD is represented by 
\begin{equation}
\langle \delta x_t \rangle \sim \frac{L }{ \langle T \rangle} t.
\end{equation}
This formula is applicable for any diffusion processes when the landscape is  periodic \cite{Reimann2001, *Reimann2002}. 
Applying the MFPT of the biased QTM, i.e., Eq.~(\ref{mfpt_qtm}), we have 
\begin{equation}
\langle \delta x_t \rangle \sim L \langle n_t \rangle \sim \frac{\varepsilon}{\mu_i}t \quad (t\to\infty),
\label{md-asympt}
\end{equation}
where $\varepsilon = p-q$.

\begin{figure*}
\includegraphics[width=.9\linewidth, angle=0]{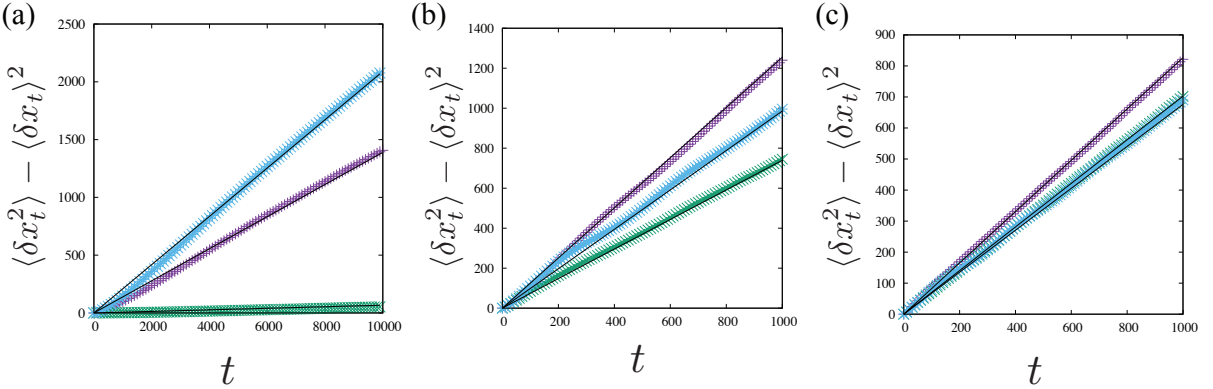}
\caption{Variance of the displacement (VD) for different $\alpha$ (a) $\alpha=0.5$, (b) $\alpha=1.5$, and (c) $\alpha=2.5$  ($p=0.8$ and $\tau_c=1$). 
Symbols are the results of VDs for three different disorder realizations ($L=10^2$). 
Solid lines are the asymptotic results, i.e., Eq.~(\ref{var_xt}). 
 }
\label{vd-sample-fluctuation}
\end{figure*}

Using Eq.~(\ref{xt_nt}), we have
\begin{equation}
\delta x_t^2 = L^2 n_t^2 + 2LC_L  n_t + C_L^2.
\end{equation}
It follows that the variance of $\delta x_t$ is given by
\begin{equation}
\langle \delta x_t^2 \rangle -\langle \delta x_t\rangle^2  \sim L^2 (\langle n_t^2 \rangle -\langle n_t\rangle^2)\quad (t\to\infty), 
\end{equation}
where we used a fact that $C_L$ and $n_t$ are independent, i.e., $\langle C_L n_t \rangle = \langle C_L \rangle \langle n_t \rangle $.
By the renewal theory \cite{Cox}, the variance of $n_t$  is derived for $t\gg 1$ as 
\begin{equation}
\langle n_t^2 \rangle -\langle n_t\rangle^2 \sim \frac{\langle T^2 \rangle - \langle T \rangle^2}{\langle T \rangle^3} t,
\end{equation} 
where $\langle T^2 \rangle$ is the second moment of the FPT. 
Therefore, the variance of $\delta x_t$ can be represented by 
\begin{equation}
\langle \delta x_t^2 \rangle -\langle \delta x_t \rangle^2 \sim 
 \left(  \frac{ \varepsilon \sigma_i^2}{   \mu_i^3} + \frac{1}{\mu_i} \right) t
\label{var_xt}
\end{equation} 
for $t\to\infty$. Figure~\ref{vd-sample-fluctuation} shows a good agreement in between numerical simulations and the theory.

\begin{figure*}
\includegraphics[width=.9\linewidth, angle=0]{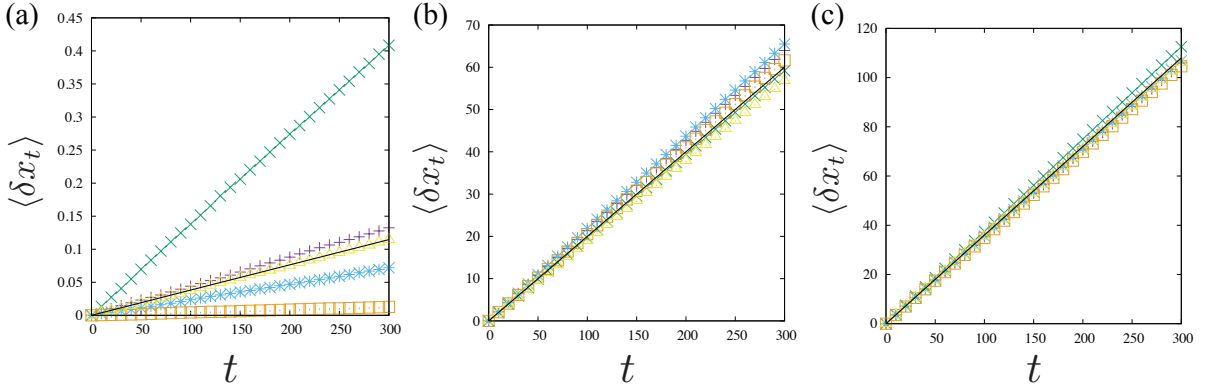}
\caption{Mean displacement (MD) for different $\alpha$ (a) $\alpha=0.5$, (b) $\alpha=1.5$, and (c) $\alpha=2.5$  ($p=0.8$ and $\tau_c=1$). 
Symbols are the results of MDs for five different disorder realizations  ($L=10^3$). We note that the average is taken only for the thermal average. Thus, 
sample-to-sample fluctuations are significant for (a).  
Solid lines are the disorder averages of the MD, i.e., $ \langle \lambda \rangle_{\rm dis} t$, where 
$\langle \lambda \rangle_{\rm dis}$ can be calculated by  Eq.~(\ref{drift_dis_ave}) for $\alpha \leq 1$ and 
$\langle \lambda \rangle_{\rm dis}=\varepsilon /\langle \mu \rangle_{\rm dis}$ for $\alpha>1$, which can be calculated by Eq.~(\ref{power-law-pdf}). 
 }
\label{md-sample-fluctuation}
\end{figure*}

\section{Self-averaging properties for drift and diffusivity}
\subsection{Drift}
The asymptotic behavior of the MD is given by Eq.~(\ref{md-asympt}). The result becomes exact for any $t>0$ when 
the initial condition is the steady state of the system, i.e., Eq.~(\ref{steady_state}), because the MD can be described as 
\begin{equation}
\langle \delta x_t  \rangle_{\rm st} = \varepsilon \langle N_t \rangle_{\rm st}
\label{mean_displacement}
\end{equation}
for any $t>0$, where  $N_t$ is the mean number of steps of a particle until time $t$ 
and $\langle \cdot \rangle_{\rm st}$ is the average when the initial condition is the steady state. 
When the initial condition is the steady state, $\langle N_t \rangle_{\rm st}$ increases linearly with time:
\begin{equation}
\langle N_t \rangle_{\rm st} = \frac{t}{\mu_i}
\label{renewal_func}
\end{equation}
for any $t>0$. Hence, the drift defined as $\lambda \equiv \langle \delta x_t  \rangle_{\rm st} /t$ is given by
\begin{equation}
\lambda_i (L) = \frac{\varepsilon}{\mu_i}
\label{drift}
\end{equation} 
for some disorder realization. 

Now, we consider sample-to-sample fluctuations of the drift. 
When the mean trapping time, $\langle \tau \rangle \equiv \int_0^\infty  \tau \psi_\alpha (\tau) d\tau$, is finite ($\alpha > 1$),
we have $\mu_i \rightarrow  \langle \tau \rangle$  ($L\to \infty$) by the law of large numbers. Therefore, 
 in the large-$L$ limit, the drift does not depend on the disorder realization (see Fig.~\ref{md-sample-fluctuation}). 
 Hence, the drift is self-averaging (SA) for $\alpha > 1$
 \cite{bouchaud90}. To quantify the SA property of $\lambda_i$, we consider the SA parameter defined as \cite{Akimoto2016, Akimoto2018}
\begin{equation}
 {\rm SA}(L;\lambda) 
\equiv \frac{\langle \lambda_i (L)^2  \rangle_{\rm dis}  - \langle  \lambda_i (L)\rangle_{\rm dis}^2}
{\langle  \lambda_i (L) \rangle_{\rm dis}^2 },
\end{equation}
where $\langle \cdot \rangle_{\rm dis}$ means the disorder average, i.e., the average obtained under different disorder realizations.
The SA parameter becomes zero in the large-$L$ limit when the drift is SA. 
Using Eq.~(\ref{drift}), we have
\begin{equation}
{\rm SA}(L;\lambda) =  \frac{\langle 1/\mu_i^2 \rangle_{\rm dis}  - \langle 1/\mu_i \rangle_{\rm dis}^2}{\langle 1/\mu_i \rangle_{\rm dis}^2 }, 
\label{SA_drift}
\end{equation} 
which is the same as the SA parameter for the diffusion coefficient in the absence of bias \cite{Akimoto2016, Akimoto2018}. 
For $\alpha>2$, the SA parameter decays as $L^{-1}$ by the central limit theorem. For $1<\alpha<2$, it goes to zero as $L\to\infty$ 
by the law large numbers. However, as shown in Appendix.~A, 
it decays non-trivially as $L^{-\alpha +1}$ in the large-$L$ limit. Therefore, sample-to-sample fluctuations remain large even for 
a relatively large $L$ when $\alpha$ closes to one.

When the mean trapping time diverges ($\alpha \leq 1$),
the law of large numbers does not hold. However, the generalized central limit theorem is still valid, which states that  
 the PDF of the normalized sum of $\tau_{n}^{(i)}$ 
follows the one-sided L\'evy distribution \cite{Feller1971}:
\begin{equation}
\dfrac{\sum_{n=1}^L \tau_{n}^{(i)}}{L^{1/\alpha}} \Rightarrow X_\alpha\quad (L\to \infty),
\label{gclt_tau}
\end{equation}
where $X_\alpha$ is a random variable following the one-sided L\'evy distribution of index $\alpha$. 
Therefore, there are large sample-to-sample fluctuations in sample mean $\mu_i$. 
The PDF of $X_\alpha$, denoted by 
$l_\alpha(x)$ with $x>0$, can be expressed as an infinite series \cite{Feller1971}
\begin{equation}
l_\alpha(x) = -\frac{1}{\pi x} \sum_{k=1}^\infty \frac{\Gamma(k\alpha +1)}{k!} (-cx^{-\alpha})^k \sin (k\pi \alpha),
\label{levy_pdf}
\end{equation}
where $c$ is a scale parameter, given by $c=\Gamma (1-\alpha) \tau_c^\alpha$ for $\psi_\alpha (\tau)$. 
Here, we define the inverse L\'evy distribution as the PDF of $X_\alpha^{-1}$:
\begin{equation}
g_\alpha(y) = -\frac{1}{\pi y} \sum_{k=1}^\infty \frac{\Gamma(k\alpha +1)}{k!} (-cy^{\alpha})^k \sin (k\pi \alpha),
\label{PDF_inverse_Levy}
\end{equation}
where the first and the second  moments of $X^{-1}_\alpha$ are given by \cite{Akimoto2016}
\begin{equation}
\langle X_\alpha^{-1} \rangle = \frac{\Gamma(\frac{1}{\alpha})}{\alpha c^{\frac{1}{\alpha}}},
\quad \langle X_\alpha^{-2} \rangle = \frac{\Gamma(\frac{2}{\alpha})}{\alpha c^{\frac{2}{\alpha}}}.
\label{moments_inv_Levy}
\end{equation}
Drift can be represented by
\begin{equation}
\lambda_i (L) = \varepsilon \frac{L}{L^{1/\alpha}}\frac{L^{1/\alpha}}{\tau_1 + \cdots + \tau_L} \sim \varepsilon L^{1-1/\alpha} X_\alpha^{-1}
\end{equation}
for $L\to \infty$. Thus, 
the PDF of $\lambda_i$ 
is described by the inverse L\'evy distribution. Therefore, 
$\lambda_i$ depends crucially on the sample of the disorder realization. 
Using the first moment of the inverse L\'evy distribution \cite{Akimoto2016, Akimoto2018}, we obtain the exact asymptotic behavior of 
the disorder average of the drift:
\begin{equation}
\langle \lambda (L)\rangle_{\rm dis} \sim \frac{\varepsilon L^{1-1/\alpha}\Gamma(\alpha^{-1})}{\alpha \tau_c \Gamma(1-\alpha)^{1/\alpha}}. 
\label{drift_dis_ave}
\end{equation}

Using the first and the second moment of $1/\mu_i$, we have the SA parameter for drift 
\begin{equation}
\lim_{L\to \infty} {\rm SA}(L; \lambda) 
=
\left\{
\begin{array}{ll}
0 &(\alpha >1)\\
\\
\dfrac{\alpha \Gamma(\frac{2}{\alpha}) }{\Gamma(\frac{1}{\alpha})^2} -1\quad &(\alpha \leq 1).
\end{array}
\right.
\label{SA_theory}
\end{equation}
For $\alpha<1$, the SA parameter is a non-zero constant, and thus $\lambda_i$ becomes non-SA; i.e., there are large sample-to-sample fluctuations in the drift [see Fig.~\ref{md-sample-fluctuation}(a)]. Therefore, the transition temperature from SA to 
non-SA behavior in the drift is given by $T_c = T_g$. 

\subsection{Diffusivity}

Here, we consider the VD to characterize the diffusivity of the system, which is 
defined as
\begin{equation}
{\rm Var} (\delta{x}_t)_{\rm st} \equiv \langle \delta{x}_t^2 \rangle_{\rm st} - \langle \delta{x}_t \rangle_{\rm st}^2.
\end{equation}
By Eq.~(\ref{var_xt}), the asymptotic behavior of the VD increases linearly with time:
\begin{equation}
{\rm Var} (\delta{x}_t)_{\rm st} \sim  \left(  \frac{ \varepsilon \sigma_i^2}{   \mu_i^3} + \frac{1}{ \mu_i} \right) t
\label{var_xt2}
\end{equation}
for $t \gg 1$ and $L\gg 1$. 
Therefore, the diffusion coefficient of the system, i.e., 
$D_i (L) \equiv \displaystyle \lim_{t\to\infty} {\rm Var} (\delta{x}_t)_{\rm st}/(2t)$, 
is given by
\begin{equation}
D_i (L) \sim   \frac{ 1}{2 }  \left(  \frac{ \varepsilon \sigma_i^2}{   \mu_i^3} + \frac{1}{ \mu_i} \right) 
\label{Di}
\end{equation}
for $L\to \infty$. 
The disorder average of $D_i (L)$ is given by 
\begin{equation}
\langle D(L) \rangle_{\rm dis} \sim   \frac{ 1}{2 }  \left(  \varepsilon \left\langle \frac{  \sigma_i^2}{   \mu_i^3} \right\rangle_{\rm dis}+ 
 \left\langle \frac{1}{ \mu_i } \right\rangle_{\rm dis} \right) 
\label{dis-Di}
\end{equation}
for $L\to \infty$. 
For $\alpha >2$, the second moment of trapping times exists; i.e., $\langle \tau^2 \rangle \equiv \int_0^\infty \tau^2 \psi (\tau)d\tau<\infty$. 
It follows that the disorder average of $D_i (L)$ is finite and given by 
\begin{equation}
\langle D (L)\rangle_{\rm dis} \sim   \frac{ 1}{2 }  \left(  \varepsilon \frac{  \langle \tau^2 \rangle - \langle \tau\rangle^2 }{   \langle \tau \rangle^3} +   
\frac{1}{ \langle \tau \rangle  }  \right) 
\label{dis-Di}
\end{equation}
 for $L\to \infty$ and $\alpha >2$. 
 
\begin{figure}
\includegraphics[width=.8\linewidth, angle=0]{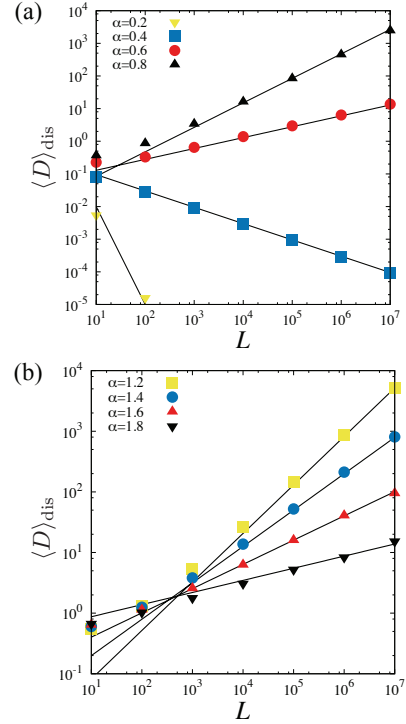}
\caption{Disorder average of the diffusion coefficient as a function of $L$ 
for different $\alpha$  ($p=0.8$ and $\tau_c=1$). 
Symbols are the results of numerical simulations. 
Solid lines are the asymptotic results, i.e., Eqs.~(\ref{scaling-D1}) and (\ref{scaling-D}). 
 }
\label{dis-ave-diffusivity}
\end{figure}

 For $\alpha <2$, the disorder average of $\{\tau_n^{(i)}\}^2 $ diverges. 
 To compute the disorder average of $\sigma_i^2/\mu_i^3$, 
we consider the scaling of the sum of $\{\tau_n^{(i)}\}^2 $.  
The PDF of $\{\tau_n^{(i)}\}^2 $ is given by
\begin{equation}
\psi_{2,\alpha}(x) = \frac{1}{2\sqrt{x}} \psi_\alpha (\sqrt{x}),
\end{equation}
because $\Pr (\tau^2 \leq x) = \Pr (\tau \leq \sqrt{x})= \int_0^{\sqrt{x}} \psi_\alpha (x')dx'$.
For $x\to \infty$, the PDF becomes
\begin{equation}
\psi_{2,\alpha}(x)  \propto x^{-1-\alpha/2} \propto \psi_{\alpha/2}(x).
\end{equation}
By the generalized central limit theorem \cite{Feller1971}, normalized sum $X_{\frac{\alpha}{2}} (L)$, defined by
\begin{equation}
X_{\frac{\alpha}{2}} (L) \equiv \frac{1}{L^{2/\alpha}}\sum_{n=1}^L \{\tau_n^{(i)}\}^2  ,
\label{gclt_2}
\end{equation}
converges in distribution to a random variable with the one-sided L\'evy distribution of index $\alpha/2$. 
For $1<\alpha <2$, $\langle \tau \rangle$ is finite but $\langle \tau^2 \rangle$ diverges. Therefore, $D_i(L)$ is proportional to 
$\sigma_i^2/\mu^3_i$, which depends on $L$. 
As shown in Appendix~B, the disorder average of $D_i (L)$ increases with $L$: 
\begin{equation}
\langle D (L)\rangle_{\rm dis} \propto L^{2-\alpha }
\label{scaling-D1}
\end{equation}
 for $L\to \infty$, which means that 
the diffusion coefficient diverges in the large-$L$ limit [see Fig.~\ref{dis-ave-diffusivity}(b)]. 
This divergence of the diffusion coefficient is a manifestation of field-induced superdiffusion in the corresponding 
annealed system, i.e., the biased CTRW \cite{Burioni2014}. This is because  
the variance of the displacement exhibits superdiffusion, i.e., ${\rm Var} (\delta{x}_t)_{\rm st} \propto t^{3-\alpha}$, which means 
that a standard diffusion coefficient diverges.

For $\alpha <1$, both the first and the second moments of the trapping times diverge. 
The scalings of the sums of $\tau_n^{(i)}$ and $\{\tau_n^{(i)}\}^2$ follows 
\begin{equation}
\sum_{n=1}^L \tau_{n} =O(L^{1/\alpha})\quad{\rm and}\quad \sum_{n=1}^L \{\tau_n^{(i)}\}^2  =O(L^{2/\alpha})
\end{equation}
for $L\to\infty$. Since $D_i (L)$ is proportional to $\sigma_i^2/\mu^3_i$, we have 
\begin{equation}
D_i (L) \propto \frac{\sigma_i^2}{\mu^3_i} \propto L^{2} \frac{\sum_{n=1}^L \tau_{n}^2}{(\sum_{n=1}^L \tau_{n})^3} 
\propto L^{2-1/\alpha}.
\label{scaling-D0}
\end{equation}
It follows that the scaling of the disorder average of $D_i (L)$ becomes 
\begin{equation}
\langle D (L) \rangle_{\rm dis} \propto L^{2-1/\alpha}
\label{scaling-D}
\end{equation}
Hence, the diffusion coefficient diverges for $\alpha > 1/2$, whereas it becomes zero 
for $\alpha <1/2$ [see Fig.~\ref{dis-ave-diffusivity}(a)]. This is physically reasonable because the variance of the displacement in the CTRW with drift becomes 
superdiffusive and subdiffusive for $\alpha >1/2$ and $\alpha <1/2$, respectively.

Let us consider the SA property for the diffusion coefficient. The SA parameter is defined as
\begin{equation}
 {\rm SA}(L;D) 
\equiv \frac{\langle D (L)^2  \rangle_{\rm dis}  - \langle D (L) \rangle_{\rm dis}^2}
{\langle D (L) \rangle_{\rm dis}^2 }.
\end{equation}
The SA parameter goes to zero in the large-$L$ limit when the diffusion coefficient is SA.

For $\alpha >2$, the second moment of trapping times exists; i.e., $\langle \tau^2 \rangle \equiv \int_0^\infty \tau^2 \psi (\tau)d\tau<\infty$. 
Therefore, sample variance $\sigma_i^2$
converges to $\langle \tau^2 \rangle -\langle \tau \rangle^2 $ as $L\to \infty$.
Hence, the sample mean and the sample mean of the squared trapping times are converges to constants, which means 
$ \langle D (L)^2  \rangle_{\rm dis}  - \langle D (L)\rangle_{\rm dis}^2 \to 0$ for $L\to \infty$.  Therefore, 
the diffusion coefficient is SA for $\alpha >2$.

For $1<\alpha<2$, the second moment of $D_i (L)$ is also calculated in Appendix.~B. As shown in Fig.~\ref{SA-diffusivity}(b), 
the SA parameter increases with $L$. In particular, it diverges as 
\begin{equation}
{\rm SA}(L;D) \propto \frac{\langle D (L)^2 \rangle_{\rm dis} }{\langle D (L) \rangle_{\rm dis}^2 }  \propto L^{ \alpha-1} 
\label{SA-scaling}
\end{equation}
for $L\to \infty$. 
It follows that  the diffusion coefficient is non-SA for $1<\alpha <2$. 

For $\alpha <1$, both the first and the second moment of the trapping times diverge. 
By Eq.~(\ref{scaling-D0}) and $\sum_{n=1}^L \tau_{n}^2 
< \left(\sum_{n=1}^L \tau_{n} \right)^3$, $D_i (L)$ can be represented as 
\begin{equation}
D_i (L) \sim \frac{\varepsilon L^{2-1/\alpha}}{2} C_i(L) ,
\end{equation}
where $C_i(L)= L^{1/\alpha} \sum_{n=1}^L \tau_{n}^2 / \left(\sum_{n=1}^L \tau_{n} \right)^3$ is a random variable depending on the disorder realization. 
Therefore, the SA parameter becomes 
\begin{equation}
 {\rm SA}(L;D) = \frac{\langle D (L)^2 \rangle_{\rm dis} }{\langle D(L) \rangle_{\rm dis}^2} -1
= \frac{\langle C(L)^2 \rangle_{\rm dis} }{\langle C (L) \rangle_{\rm dis}^2} -1,
\end{equation}
which is a finite value because 
$1/ \left(\sum_{n=1}^L \tau_{n} \right)^3 < C_i(L)  <1$, i.e., 
$0< \langle C_i(L) \rangle_{\rm dis} <1$ and $0< \langle C_i(L)^2 \rangle_{\rm dis} <1$.  Thus, 
${\rm SA}(L;D) \to S(\alpha) >0$ for $L\to \infty$; i.e., 
the diffusion coefficient is non-SA for $\alpha <1$ [see Fig.~\ref{SA-diffusivity}(a)]. 

\begin{figure}
\includegraphics[width=.8\linewidth, angle=0]{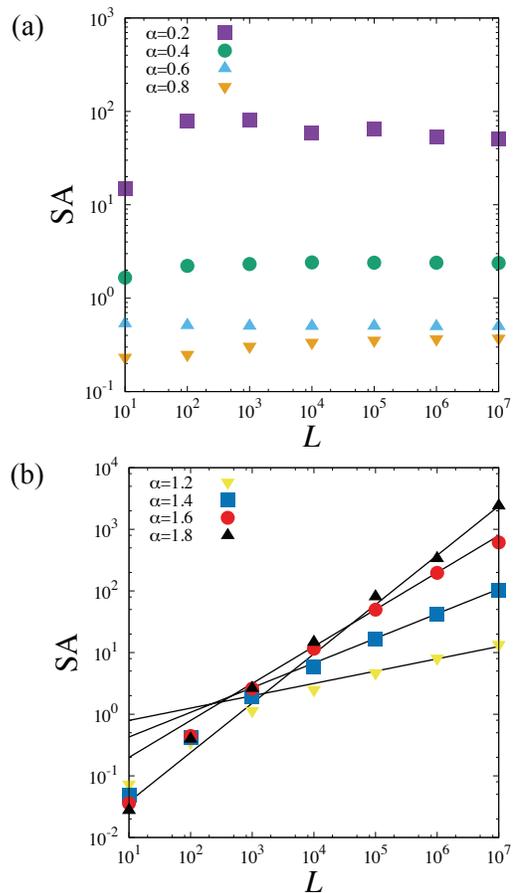}
\caption{Self-averaging parameter of the diffusion coefficient as a function of $L$ 
for different $\alpha$   ($p=0.8$ and $\tau_c=1$). 
Symbols are the results of numerical simulations. 
Solid lines are the asymptotic results, i.e., Eq.~(\ref{SA-scaling}). 
 }
\label{SA-diffusivity}
\end{figure}

\section{Conclusion}

\begin{figure}
\includegraphics[width=.9\linewidth, angle=0]{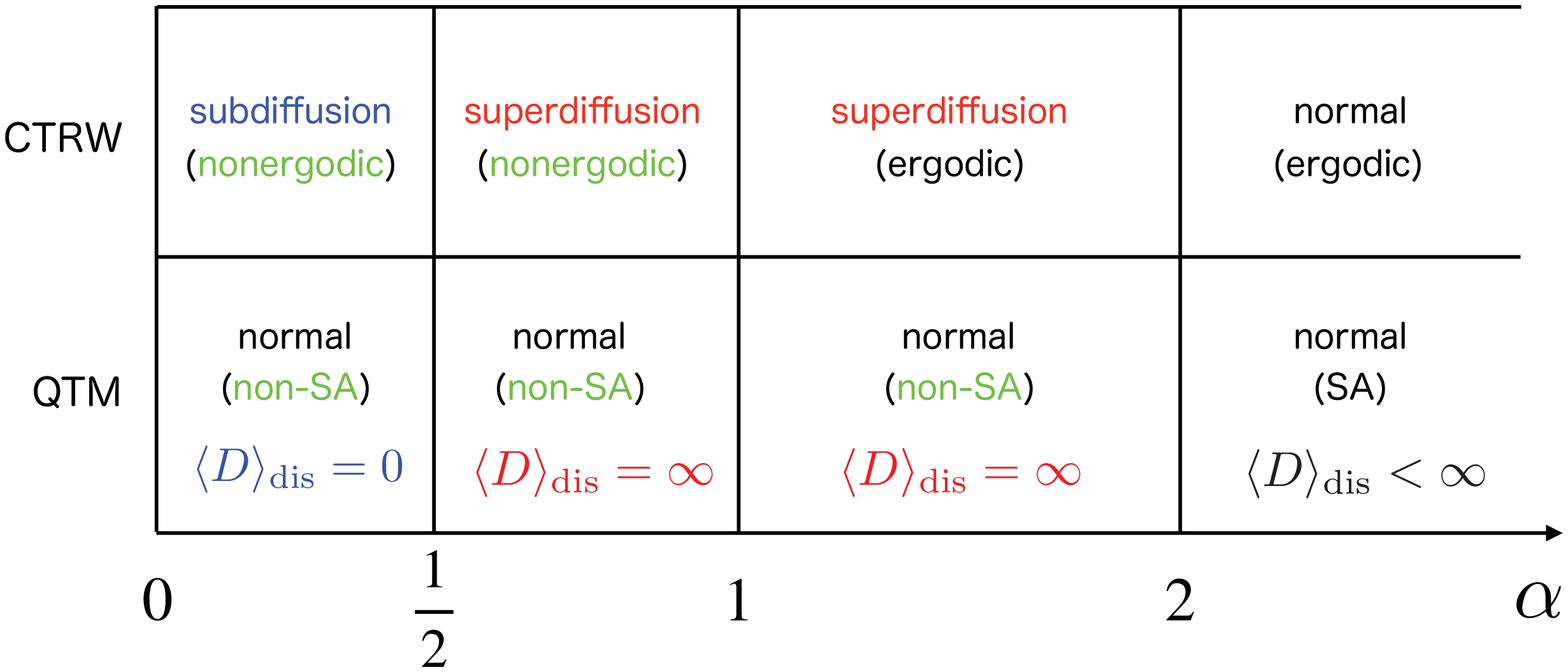}
\caption{Phase diagram based on diffusivity in the CTRW and QTM. 
Divergence of the disorder average of the diffusion coefficient in the large-$L$ limit implies superdiffusion in the CTRW. 
On the other hand, zero of  the disorder average of the diffusion coefficient 
implies subdiffusion in the CTRW. Note that ergodicity in the CTRW does 
not imply SA in the corresponding quenched model. 
 }
\label{phase}
\end{figure}

The MD and the VD are always normal in a biased QTM with a finite system size with the aid of a stationary steady state, 
which is different from the biased CTRW.
Using the FPT statistics, we have provided exact results for the drift and the diffusion coefficient in the biased QTM. 
We have found that anomaly of the disorder average of the diffusion coefficient 
is a manifestation of anomalous diffusion in the corresponding annealed model (CTRW). In particular, 
 divergence and zero of the disorder average of the diffusion coefficient, i.e., $\langle D \rangle_{\rm dis}$, 
 in the biased QTM implies superdiffusion and subdiffusion in the biased CTRW, respectively (see Fig.~\ref{phase}). 
 Moreover, we have introduced the SA parameter to quantify the SA property. Transition points between SA and non-SA 
are $\alpha=1$ and $\alpha=2$ for the MD and the VD, respectively. 

\section*{Acknowledgement}
T.A. was supported by JSPS Grant-in-Aid for Scientific Research (No. C JP18K03468) and  
K.S. was supported by JSPS Grants-inAid for Scientific Research (JP17K05587).


\appendix

\section{Scaling of the SA parameter for drift}
Here, we show a power-law decay of the SA parameter for drift. For $1<\alpha <2$, it decays as $L^{1-\alpha}$, where $L$ is the system size. 
Drift for a sample realization is given by 
\begin{equation}
\lambda (L) = \frac{\varepsilon L}{\tau_1 + \cdots + \tau_L} = \frac{\varepsilon L}{ L^{1/\alpha} \tilde{X}_\alpha (L) + \langle \tau \rangle L },
\end{equation}
where 
\begin{equation}
\tilde{X}_\alpha (L)  = \frac{\tau_1 + \cdots + \tau_L - \langle \tau \rangle L }{ L^{1/\alpha} }.
\end{equation}
In the large-$L$ limit, the PDF of $\tilde{X}_\alpha (L)$ converges to the L\'evy distribution of index $\alpha$. 
For $L^{1/\alpha -1} \tilde{X}_\alpha (L) \ll \langle \tau \rangle \varepsilon $, $\lambda (L)$ 
becomes 
\begin{equation}
\lambda (L) \cong \frac{\varepsilon}{\langle \tau \rangle} - \frac{\varepsilon L^{1/\alpha-1} \tilde{X}_\alpha (L) } {\langle \tau \rangle^2 }.
\end{equation}
On the other hand, it becomes 
\begin{equation}
\lambda (L) \cong  \frac{\varepsilon L^{1-1/\alpha}  } {\tilde{X}_\alpha (L) }
\end{equation}
for $L^{1/\alpha -1} \tilde{X}_\alpha (L) \gg \langle \tau \rangle \varepsilon $.
The ensemble average of $\tilde{X}_\alpha (L)^2$ restricted in $\tilde{X}_\alpha (L) <  L^{1 - 1/\alpha } \langle \tau \rangle \varepsilon $, denoted by $\langle \tilde{X}_\alpha (L)^2 \rangle_{<L^{1 - 1/\alpha }}$, is given by 
\begin{eqnarray}
\langle \tilde{X}_\alpha (L)^2 \rangle_{<L^{1 - 1/\alpha }} &\cong& \int_0^{L^{1 - 1/\alpha }  \langle \tau \rangle \varepsilon} x^2 \psi_\alpha (x) dx \\
&\propto& L^{3- \frac{2}{\alpha} - \alpha}.
\end{eqnarray}
Moreover, the ensemble average of $1/\tilde{X}_\alpha (L)^2$ restricted in $\tilde{X}_\alpha (L) >  L^{1 - 1/\alpha } \langle \tau \rangle \varepsilon $, denoted by $\langle 1/\tilde{X}_\alpha (L)^2 \rangle_{>L^{1 - 1/\alpha }}$, is given by 
\begin{eqnarray}
\langle \tilde{X}_\alpha (L)^{-2} \rangle_{>L^{1 - 1/\alpha }} &\cong& \int_{L^{1 - 1/\alpha }\langle \tau \rangle \varepsilon}^\infty   
x^{-2} \psi_\alpha (x) dx \\
&\propto& L^{-(2+\alpha)(1 -\frac{1}{\alpha})}.
\end{eqnarray}
The variance of $\lambda (L)$ becomes
\begin{eqnarray}
\langle \lambda(L)^2 \rangle_{\rm dis} - \langle \lambda(L) \rangle_{\rm dis}^2 &&\propto L^{\frac{2}{\alpha}-2} \langle \tilde{X}_\alpha (L)^2 \rangle_{<L^{1 - 1/\alpha }} \nonumber\\
+&& L^{2-\frac{2}{\alpha}} \langle \tilde{X}_\alpha (L)^{-2} \rangle_{>L^{1 - 1/\alpha }}. 
\end{eqnarray}
It follows that the scaling of the SA parameter for drift becomes 
\begin{equation}
 {\rm SA}(L; \lambda) \propto L^{1-\alpha}
\end{equation}
for $L\to \infty$ and $1<\alpha <2$. 

\section{Scaling for the SA parameter for diffusivity}

For $1<\alpha <2$, the diffusion coefficient is proportional to $\sigma_i^2/\mu_i^3$, which can be written as
\begin{equation}
\dfrac{\sigma_i^2}{\mu_i^3} = \frac{(\tau_1^2 + \cdots + \tau_L^2)L^2}{\langle \tau \rangle^3 L^3 \left( 1 + 
\dfrac{L^{1/\alpha -1}}{\langle \tau \rangle} \tilde{X}_\alpha (L)\right)^3} .
\end{equation}
For $\tilde{X}_\alpha (L) \ll  L^{1 - 1/\alpha } \langle \tau \rangle$, 
\begin{equation}
\frac{\sigma_i^2}{\mu_i^3} \propto \frac{\tau_1^2 + \cdots + \tau_L^2}{ L } =
L^{\frac{\alpha}{2}-1} X_{\frac{\alpha}{2}}(L)
\label{scaling-sigma2mu-rest}
\end{equation}
in the large-$L$ limit. Moreover, variable $X_{\frac{\alpha}{2}}(L)$ satisfies 
\begin{equation}
X_{\frac{\alpha}{2}}(L) < \frac{ (\tau_1 + \cdots + \tau_L)^2}{L^{2/\alpha}} \sim \langle \tau \rangle^2 L^{2-\frac{2}{\alpha}}
\label{ineq-x}
\end{equation}
for $\tilde{X}_\alpha (L) \ll  L^{1 - 1/\alpha } \langle \tau \rangle$. 
Scaling of the disorder average of $D_i$ follows 
\begin{eqnarray}
\left\langle \frac{\sigma_i^2}{\mu_i^3} \right\rangle_{\rm dis} &\propto& 
L^{\frac{2}{\alpha}-1} \langle X_{\frac{\alpha}{2}}(L) \rangle_{<L^{2-\frac{2}{\alpha}}} \nonumber\\
&&+ L^{2- \frac{3}{\alpha}} \left\langle \frac{X_{\frac{\alpha}{2}}(L)}{\tilde{X}_\alpha (L)^3} \right\rangle_{>L^{2-\frac{2}{\alpha}}}, 
\end{eqnarray}
where $\langle \cdot \rangle_{<L^{2-\frac{2}{\alpha}}}$ and $\langle \cdot \rangle_{>L^{2-\frac{2}{\alpha}}}$ represent 
 the ensemble averages restricted in $X_{\frac{\alpha}{2}}(L) <L^{2-\frac{2}{\alpha}}$ and $X_{\frac{\alpha}{2}}(L) >L^{2-\frac{2}{\alpha}}$, 
 respectively. Using Eq.~(\ref{ineq-x}), we obtain $L^{\frac{2}{\alpha}-1} \langle X_{\frac{\alpha}{2}}(L) \rangle_{<L^{2-\frac{2}{\alpha}}} 
 \propto L^{2-\alpha}$. 
 The second term can be evaluated as  
 \begin{equation}
 \left\langle \frac{X_{\frac{\alpha}{2}}(L)}{\tilde{X}_\alpha (L)^3} \right\rangle_{>L^{2-\frac{2}{\alpha}}}
 < \left\langle \frac{1}{\tilde{X}_\alpha (L)} \right\rangle_{>L^{1-\frac{1}{\alpha}}}.
 \end{equation}
 Thus, the second term can be neglected because the order is smaller than the first one, i.e., $L^{2-\alpha}$. 
 It follows that the scaling of the disorder average of $D_i$ becomes 
\begin{equation}
\langle D (L) \rangle_{\rm dis}  \propto \left\langle \frac{\sigma_i^2}{\mu_i^3} \right\rangle_{\rm dis} \propto L^{2-\alpha} .
\end{equation}
By Eq.~(\ref{scaling-sigma2mu-rest}), one obtains the second moment of $D_i(L)$ in a similar way. It becomes 
\begin{equation}
\langle D (L)^2 \rangle_{\rm dis}  \propto L^{ 2\left(\frac{2}{\alpha}-1 \right)} \langle X_{\frac{\alpha}{2}}(L)^2 \rangle_{L^{2-\frac{2}{\alpha}}}
\end{equation}
for $L\to \infty$. 
Therefore, the SA parameter for the diffusion coefficient is given by  
\begin{equation}
{\rm SA}(L; D) \propto \frac{\langle D (L)^2 \rangle_{\rm dis} }{\langle D (L) \rangle_{\rm dis}^2 }  \propto L^{ \alpha-1} 
\end{equation}
for $L\to \infty$ and $1<\alpha <2$. 

%

\end{document}